\documentclass[preprint,superscriptaddress,showpacs,nofootinbib,amsmath, amssymb, floatfix, author-numerical]{revtex4-1}

\usepackage{color}
\usepackage[dvips]{graphicx}

\usepackage[colorinlistoftodos, textwidth=4cm, shadow]{todonotes}

\begin{document}

\title{Using entropy measures for comparison of software traces}


\author{A. V. Miranskyy}
\email{amiransk@alumni.uwo.ca}
\affiliation{Department of Applied Mathematics, University of Western Ontario, London, Ontario N6A 5B7, Canada}
\affiliation{IBM Toronto Software Lab, Markham, Ontario L6G 1C7, Canada}

\author{M. Davison}
\email{mdavison@uwo.ca}
\affiliation{Department of Applied Mathematics, University of Western Ontario, London, Ontario N6A 5B7, Canada}
\affiliation{Department of Statistical \& Actuarial Sciences, University of Western Ontario, London, Ontario N6A 5B7, Canada}
\affiliation{Richard Ivey School of Business, University of Western Ontario, London, Ontario N6A 5B7, Canada}

\author{M. Reesor}
\email{mreesor@uwo.ca}
\affiliation{Department of Applied Mathematics, University of Western Ontario, London, Ontario N6A 5B7, Canada}
\affiliation{Department of Statistical \& Actuarial Sciences, University of Western Ontario, London, Ontario N6A 5B7, Canada}

\author{S. S. Murtaza}
\email{smurtaza@alumni.uwo.ca}
\affiliation{Department of Computer Science, University of Western Ontario, London, Ontario N6A 5B7, Canada}

\begin{abstract}
The analysis of execution paths (also known as software traces)
collected from a given software product can help in a number of areas
including software testing, software maintenance and program
comprehension. The lack of a scalable matching algorithm operating on detailed execution paths motivates the search for an alternative solution.

This paper proposes the use of word entropies for the classification of software traces.  Using a well-studied defective software as an example, we investigate the application of both Shannon and extended entropies (Landsberg-Vedral, R\'{e}nyi and Tsallis) to the classification of traces related to various software defects.  Our study shows that using entropy measures for comparisons gives an efficient and scalable method for comparing traces.  The three extended entropies, with parameters chosen to emphasize rare events, all perform similarly and are superior to the Shannon entropy.
\end{abstract}



\maketitle

\section{Introduction}\label{sec:1}
A software execution trace is a log of information
captured during a given execution run of a computer program. For example,
the trace depicted in Figure \ref{fig:trc_ex} shows the program flow entering function f1; calling f2
from f1; f2 recursively calling itself, and eventually exiting these
functions. In order to capture this information, each function in the
software is instrumented to log both: entry to it and exit from it.

\begin{figure}[htb]
\centering
\begin{minipage}[H]{1in}
	\begin{verbatim}
	1 f1 entry
	2 | f2 entry
	3 | | f2 entry
	4 | | f2 exit
	5 | f2 exit
	6 f1 exit    
	\end{verbatim}
\end{minipage}    
\caption{An example of a trace}
\label{fig:trc_ex}
\end{figure}

The comparison of program execution traces is important for a number
of problem areas in software development and use. In the area of
testing trace comparisons  can be used to: 1) determine
how well user execution paths (traces collected in the field) are
covered in testing \cite{cotroneo_investigation_2007, elbaum_trace_2007, moe_using_2002}; 2) detect anomalous behavior arising during a
component's upgrade or reuse \cite{mariani_dynamic_2007}; 3) map and classify defects \cite{haran_applying_2005, podgurski_automated_2003, yuan_automated_2006}; 4) determine redundant test cases executed by one or more test teams \cite{rothermel_empirical_1998}; and
5) prioritize test cases (to maximize execution path coverage with a
minimum number of test cases) \cite{elbaum_selecting_2004, miranskyy_usage_2007}. Trace comparisons are also used in
operational profiling (for instance, in mapping the frequency of
execution paths used by different user classes) \cite{moe_using_2002} and intrusion analysis
(e.g., detecting deviations of field execution paths from expectations) \cite{lee_learning_1997}.

For some problems such as test case prioritization,
traces gathered in a condensed form (such as a vector of executed
function names or caller-callee pairs) are adequate \cite{elbaum_selecting_2004}. However for
others, such as the detection of missing coverage and anomalous
behavior using state machines, detailed execution paths are necessary \cite{cotroneo_investigation_2007, mariani_dynamic_2007}.
The time required for analyzing traces can be critical.  Examples of the use of trace analysis in practice include: 1) a customer support analyst may use traces to map a reported defect onto an existing set of defects in order to identify the problem's root cause; 2) a development analyst working with the testing team may use trace analysis to identify missing test coverage that resulted in a user-observed defect.

\textit{Research Problem and Practical Motivation}: To be compared and analyzed, traces must be converted into an abstract format.  Existing work has progressed by representing traces as signals \cite{kuhn_exploiting_2006},  finite automata \cite{mariani_dynamic_2007}, and complex networks \cite{cai_software_2009}. Unfortunately, many trace comparison techniques are not scalable \cite{miranskyy_sift:_2008, davison_patent_2006}. For example the finite-state automata based kTail algorithm, when applied to representative traces, did not terminate even after 24 hours of execution \cite{ cotroneo_investigation_2007}; similar issues have been experienced with another finite-state automata algorithm, kBehavior \cite{miranskyy_sift:_2008}. These observations highlight the need for fast trace matching solutions.

Based on our experience, support personnel of a large-scale industrial
application with hundreds of thousands of installations can collect
tens of thousands of traces per year. Moreover, a single trace collected on a
production system is populated at a rate of millions of records per
minute. Thus, there is a clear need for scalable trace comparison techniques.

\textit{Solution Approach}: The need to compare traces, together with a
lack of reliable and scalable tools for doing this, motivated us to
investigate alternate solutions. To speed up trace comparisons, we
propose that a given set of traces first be filtered, rejecting
those that are not going to match with the test cases, allowing just
the remaining few to be compared for target purposes. The underlying
assumption (based on our practical experience) is that most traces are
very different, just a few are even similar, and only
a very few are identical.

This strategy is implemented and validated in the
\textbf{S}calable \textbf{I}terative-un\textbf{F}olding
\textbf{T}echnique (SIFT) \cite{miranskyy_sift:_2008}. The collected traces are first compressed
into several levels prior to comparison. Each level of compression
uses a unique signature or
``fingerprint''\footnote{ The
fingerprint of the next iteration always contains more information than
the fingerprint of the previous iteration, hence the term \textit{unfolding}. }.
Starting with the highest compression level, the traces are
compared, and unmatched ones are rejected. Iterating through the lower
levels until the comparison process is complete leaves only traces that
match at the lowest (uncompressed) level. The SIFT objective ends
here. The traces so matched can then be passed on to external tools, such as the ones presented here, for
further analysis such as defect or security breach identification.

The process of creating a fingerprint can be interpreted as a map from
the very high dimensional space of traces to a low (ideally
one) dimensional space. Simple examples of such fingerprints are 1) the
total number of unique function names in a trace; and 2) the number of
elements in a trace. However, while these fingerprints may be useful
for our purposes, neither are sufficient. The ``number
of unique function names'' fingerprint does not
discriminate enough -- many quite dissimilar traces can share the same
function names called. At the other extreme, the number of elements in
the trace discriminates too much -- traces which are
essentially similar may nonetheless have varying numbers of elements. The mapping
should be such that projections of traces of different types should be
positioned far apart in the resulting small space.

Using the frequency of the function names called is the next step in
selecting useful traces. A natural one dimensional representation of
this data is the Shannon information \cite{shannon_mathematical_1948}, mathematically identical to the
entropy of statistical mechanics. Other forms of entropy/information,
obeying slightly less restrictive axiom lists, have been defined \cite{aczel_measures_1975}. These extended entropies (as reviewed in \cite{davison_extended_2005}) are indexed by a parameter
$q$ which, when $q = 1$ reduces them to the traditional Shannon entropy and which can be set to make them more ($q < 1$) or less ($q > 1$) sensitive to the frequency of rare events, improving the classification power of algorithms. Indeed, an extended R\'{e}nyi entropy \cite{renyi_probability_1970} with $q = 0$ (known as the Hartley entropy of information theory) applied in the context of this paper returns the ``number of unique function names'' fingerprint.

The entropy concept can also be extended in another way. Traces differ
not only in which functions they call but in the pattern linking the
call of one function with the call of another. As such it makes sense
to collect not only the frequency of function calls, but also the
frequency of calling given pairs, triplets, and in general
$l$-tuples of calls. The frequency information
assembled for these ``$l$-words''
can be converted into ``word entropies'' for further discriminatory power. In
addition each record in a trace can be encoded in different ways
(denoted by $c$) by incorporating various
information such as a record's function name or type.

\textit{Research Question:} In this paper we study the applicability of the Shannon  entropy \cite{shannon_mathematical_1948} and
the Landsberg-Vedral \cite{landsberg_distributions_1998}, R\'{e}nyi \cite{renyi_probability_1970}, and Tsallis \cite{tsallis_possible_1988} entropies to the comparison and classification of traces related to various software defects. We also study the effect of $q$, $l$, and $c$ values on the classification power of the entropies. 

Note that the idea of using word entropies for general classification problems is not new to this paper. Similar work has been done to apply word entropy classification
techniques to problems arising in  biology \cite{vinga_renyi_2004, dehmer_hist_2011}, chemistry \cite{dehmer_hist_2011}, analysis of natural languages \cite{ebeling_word_1992}, and image processing \cite{bhandari_some_1993}. In the context of software traces, the Shannon entropy has been used to measure trace complexity \cite{Hamou-Lhadj_measuring_2008}.  However, no one has yet applied word entropies to compare software traces.

The structure of the paper is as follows: in Section~\ref{sec:2} we define
entropies and explain the process of trace entropy calculation. The way in which entropies are used to classify traces is shown in 
Section~\ref{sec:3}. Section~\ref{sec:4} provides a case study which describes and validates the application of entropies for trace classification, and Section~\ref{sec:5} summarizes the paper.

\section{Entropies and Traces: definitions}\label{sec:2}
In this section we describe techniques for extracting the probability
of various events from traces (Section ~\ref{sec:2.1}) and the way we use this
information to calculate trace entropy (Section ~\ref{sec:2.2}).

\subsection{Extraction of probability of events from traces}\label{sec:2.1}

A trace can be represented as a string, in which each trace record is
encoded by a unique character. We concentrate on the following three character types
$c$:

\begin{enumerate}
	\item Record's function name ($F$),
	\item Record's type ($FT$),
	\item Record's function names, type, and depth in the call tree
($FTD$).
\end{enumerate}

In addition, we can generate consecutive and overlapping
substrings\footnote{ The substring can start at any character $i$, where 
$i \leq n-l+1$.} of length $l$ from a string. We call such substrings $l$-words. For example, a string ``ABCA'' contains the following 2-words: ``AB''; ``BC''; and ``CA''.

One can consider a trace as a message generated by a source with source dictionary $A = \{ {a_1},{a_2}, \ldots ,{a_n}\} $ consisting of $n$
$l$-words ${a_i}$, $i=1,2,\ldots,n$, and discrete probability distribution $P = \{ {p_1},{p_2}, \ldots ,{p_n}\} $, where ${p_i}$ is the probability ${a_i}$ is observed. To illustrate these ideas, the dictionaries $A$ and their respective probability distributions $P$ for various values of $c$ and
$l$ are summarized in Table~\ref{tab:dic_of_trc} for the trace shown in Figure~\ref{fig:trc_ex}.

\begingroup
\squeezetable
\begin{table*}[htbp]
\caption{Dictionaries of a trace given in Figure \ref{fig:trc_ex}}
\label{tab:dic_of_trc}
\begin{ruledtabular}
\begin{tabular}{lrrll}

$ C$
 & 
$ l$
 & 
$ n$
 & 
$ A$
 & 
$ P$
 \\
\hline

 $F$
 & 
 1
 & 
 2
 & 
 f1, f2
 & 
 1/3, 2/3
 \\

 $F$
 & 
 2
 & 
 3
 & 
 f1-f2, f2-f2, f2-f1
 & 
 1/5, 3/5, 1/5
 \\

 $F$
 & 
 3
 & 
 3
 & 
 f1-f2-f2, f2-f2-f2, f2-f2-f1
 & 
 1/4, 1/2, 1/4
 \\

 $FT$
 & 
 1
 & 
 4
 & 
 f1-entry, f1-exit, f2-entry, f2-exit
 & 
 1/6, 1/6, 1/3, 1/3
 \\

 $FTD$
 & 
 1
 & 
 6
 & 
 f1-entry-depth1, f1-exit-depth1, 

 f2-entry-depth2,
 & 
 1/6, 1/6, 1/6,
 \\

 & 

 & 

 & 
 f2-exit-depth2, f2-entry-depth3, f2-exit-depth3
 & 
  1/6, 1/6, 1/6
 \\

\end{tabular}
\end{ruledtabular}
\end{table*}
\endgroup

Define a function $\alpha$ that, given a trace $t$, will return a discrete probability distribution $P$ for $l$-words of length $l$ and characters of type $c$:
\begin{eqnarray}\label{eq:P}
P \leftarrow \alpha (t;l,c).
\end{eqnarray}
The above empirical probability distribution $P$ can now be used to calculate the entropy of a given trace for a specific $l$-word with type-$c$ characters. For notation convenience we suppress the dependence of $P$ (and the individual $p_{i}$'s) on the
$t$, $l$, and $c$. We now define entropies and discuss how we can utilize $P$ to compute them.

\subsection{Entropies and traces}\label{sec:2.2}

The Shannon entropy \cite{shannon_mathematical_1948} is defined as
\begin{eqnarray}\label{eq:S}
{H_S}(P) =  - \sum\limits_{i = 1}^n {{p_i}{{\log }_b}{p_i}} ,
\end{eqnarray}
where $P$ is the vector containing probabilities of  the $n$ states, and $p_{i}$ is the probability of $i$-th state. The logarithm base $b$ controls the units of entropy. In this paper we set $b=2$, to measure entropy in bits.

Three extended entropies, Landsberg-Vedral \cite{landsberg_distributions_1998}, R\'{e}nyi \cite{renyi_probability_1970}, and Tsallis \cite{tsallis_possible_1988} are defined as:
\begin{eqnarray}\label{eq:LHT}
	{H_L}(P;q) &=& \frac{{1 - 1/Q(P;q)}}{{1- q}}, \nonumber \\
	{H_R}(P;q) &=& \frac{{{{\log }_2}\left[ {Q(P;q)}\right]}}{{1 - q}}, \text{ and}\\
	{H_T}(P;q) &=& \frac{{Q(P;q) - 1}}{{1 -q}}, \nonumber
\end{eqnarray}
respectively, where $q \ge 0$ is the entropy index, and
\begin{eqnarray}
Q(P;q) = \sum\limits_{i = 1}^n {p_i^q.} 
\end{eqnarray}

These extended entropies reduce to the Shannon entropy (by L'H\^{o}pital's rule) when $q = 1$. The extended entropies are more sensitive to states with small probability of occurrence than the Shannon entropy for $0 < q < 1$. Setting $q > 1$ leads to increased sensitivity of the extended entropies to states with high probability of occurrence.

The entropy $Z$ of a trace $t$ for a given $l$, $c$, and $q$ is calculated by inserting the output of Equation~(\ref{eq:P}) into one of the entropies described in Equations (\ref{eq:S}) and (\ref{eq:LHT}):
\begin{eqnarray}
Z \leftarrow {H_E}\left[ {\alpha (t;l,c);q} \right]
\end{eqnarray}
where $E \in \{ L,R,T,S\} $. Note that if $E=S$ then $q$ is ignored, since in fact it must be that $q = 1$.

\section{Using entropies for classification of traces}\label{sec:3}
A typical scenario for trace comparison is the following. A software
service analyst receives a phone call from a customer reporting
software failure. The analyst must quickly determine the root cause
of this failure and identify if 1) this is a rediscovery of a known
defect exposed by some other customer in the past or 2) this is a newly
discovered defect. If the first case is correct then the analyst
will be able to provide the customer with a fix-patch or
describe a workaround for the problem. If the second hypothesis is
correct the analyst must alert the maintenance team and start a full
scale investigation to identify the root-cause of this new problem. In
both cases time is of the essence -- the faster the root
cause is identified, the faster the customer will receive a fix to the
problem.

In order to validate the first hypothesis, the analyst asks the
customer to reproduce the problem with a trace capturing facility
enabled. The analyst can then compare the newly collected trace against
a library of existing traces collected in the past (with known
root-causes of the problems) and identify potential candidates for
rediscovery. To identify a set of traces relating to similar
functionality the library traces are usually filtered by names of
functions present in the trace of interest. After that the filtered
subset of the library traces is examined manually to identify common
patterns with the trace of interest.

If the analyst finds an existing trace with common patterns then the trace corresponds to a rediscovered defect.  
Otherwise the analyst can conclude\footnote{
This is a simplified description of the analysis process. In practice
the analyst will examine defects with similar symptoms, consult with
her peers, search the database with descriptions of existing problems, etc.
} 
that this failure relates to a newly discovered defect. This process is similar
in nature to an Internet search engine. A user provides keywords of interest and the engine's algorithm returns a
list of web pages ranked according to their relevance. The
user examines the returned pages to identify her most relevant ones.

To automate this approach using entropies as fingerprints, we need an
algorithm that would compare a trace against a set of traces, rank this
set based on the relevance to a trace of interest, and then return the
top $X$ closest traces for manual examination to
the analyst. In order to implement this algorithm we need the measure of
distance between a pair of traces described
in Section~\ref{sec:3.1}, and the ranking algorithm described in Section~\ref{sec:3.2}. The algorithm's efficiency is analyzed in Section~\ref{sec:3.3}.

\subsection{Measure of distance between a pair of traces}\label{sec:3.1}

We can obtain multiple entropy-based fingerprints for a trace by varying values of $E$, $q$, $l$ and $c$. Denote a complete set of 4-tuples of
$[E,q,l,c]$ as $M$. We define the distance between a pair of
traces $t_{i}$ and $t_j$ as:
\begin{eqnarray}\label{eq:d_all}
D({t_i},{t_j};M) = \left( {\sum\limits_{k = 1}^m {{{\left| {\frac{
{{H_{{E_k}}}\left[ {\alpha ({t_i};{l_k},{c_k});{q_k}} \right] - {H_{{E_k}}}\left[ {\alpha ({t_j};{l_k},{c_k});{q_k}} \right]
}}{{\max \left\{ {{H_{{E_k}}}\left[ {\alpha (t;{l_k},{c_k});{q_k}} \right]} \right\}}}} \right|}^w}} } \right)^{1/w} ,
\end{eqnarray}where $m$ is the number of elements in $M$, $w$ controls\footnote{ We do not have a theoretical rationale for selecting an optimal value of $w$; experimental analysis shows that our results are robust to the value of $w$ (c.f.,  Appendix~\ref{sec:p} for more details).} the type of ``norm'', 
$\max \left\{
{{H_{{E_k}}}\left[ {\alpha (t;{l_k},{c_k});{q_k}} \right]}
\right\}$ 
denotes  the maximum value of ${H_{{E_k}}}$ for the complete set
of traces under study for a given $q_k$, $l_k$, and $c_k$; $k$ indexes the 4-tuple $[E_k,q_k,l_k,c_k]$ in $M$ with $E_k$, $q_k$, $l_k$, and $c_k$ indicating the entropy, $q$, $l$-word length and character type for the $k$-th 4-tuple, $k=1,\ldots,m$.

 This denominator is used as a normalization factor to set equal weights to fingerprints related to different 4-tuples in $M$.

Formula \ref{eq:d_all} satisfies three of the four usual conditions of a metric:
\begin{eqnarray}
D({t_i},{t_j};M) &\ge& 0, \nonumber \\ 
D({t_i},{t_j};M) &=& D({t_j},{t_i};M), \\ 
D({t_i},{t_k};M) &\le& D({t_i},{t_k};M) + D({t_k},{t_j};M). \nonumber 
\end{eqnarray}

However, the fourth metric condition, $D({t_i},{t_j};M) = 0
\Leftrightarrow {t_i} = {t_j}$ (identity of indiscernibles),
holds true only for the fingerprints of traces; the actual traces may
be different even if their entropies are the same. In other words, the
identity of indiscernibles axiom only ``half'' holds: 
${t_i} = {t_j} \Rightarrow D({t_i},{t_j};M) = 0,$ but
$D({t_i},{t_j};M) = 0\not  \Rightarrow {t_i} ={t_j}.$ As such, $D$ represents a
pseudo-metric. Note that $D({t_i},{t_j};M) \in [0,\infty )$ and our
hypothesis is the following: the larger the value of $D$, the further apart are the traces.

Note that for a single pair of entropy-based fingerprints the normalization factor in Equation~(\ref{eq:d_all}) can be omitted and we define $D$ as
\begin{eqnarray}\label{eq:d_single}
D({t_i},{t_j};&E&,q,l,c) 
=\left|
{{H_E}\left[ {\alpha ({t_i};l,c);q} \right] - {H_E}\left[ {\alpha
({t_j};l,c);q} \right]} \right|. 
\end{eqnarray}

Entropies have the drawback that they cannot differentiate dictionaries of events, since entropy
formulas operate only with probabilities of events. Therefore,
the entropies of the strings ``f1-f2-f3-f1'' and ``f4-f5-f6-f4'' will be exactly
the same for any value of $E$, $l$, $c$, and $q$. The simplest solution is to do a pre-filtering of traces in $T$ in the spirit of the SIFT framework described
in Section~\ref{sec:1}. For example, one can filter out all the traces that do not contain ``characters'' (e.g., function names) present in the trace of interest before using entropy-based fingerprints.

We now define an algorithm for ranking a set of traces with respect to
the trace of interest.

\subsection{Traces ranking algorithm}\label{sec:3.2}

Given a task of identifying the top $X$ closest classes of traces from a set of traces $T$ closest to trace $t$ we employ the following algorithm:

\begin{enumerate}
	\item Calculate the distance between $t$ and each trace in $T$;
	\item Sort the traces in $T$ by their distance to trace $t$ in ascending order;
	\item Replace the vector of sorted traces with the vector of classes (e.g.,
defect IDs) to which these traces map;
	\item Keep the first occurrence (i.e., the closest trace) of each class in
the vector and discard the rest;
	\item Calculate the ranking of classes taking into account ties using the
``modified competition ranking''
\footnote{ The ``modified
competition ranking'' assigns the same rank to items
deemed equal and leaves the ranking gap before the equally ranked items. For example, if A is ranked ahead of B and C (considered
equal), which in turn are ranked ahead of D then the ranks are:  A gets rank 1, B and C get rank 3, and D gets
rank 4.} 
approach;
	\item Return a list of classes with ranking smaller than or equal to $X$.
\end{enumerate}

The ``modified competition ranking'' can be interpreted as a worst case scenario
approach. The ordering of traces of equal ranks is arbitrary; therefore
we examine the case when the most relevant trace will always
reside at the bottom of the returned list. To be conservative, we
consider the outcome in which our method returns a trace in the top
$X$ positions as being in the $X$-th position.

Now consider an example of the algorithm.

\subsubsection{Traces ranking algorithm: example}~\label{sec:3.2.1}

Assume we have five traces $t_{i}$, $i=1..5$ related to four software defects $d_{j}$, $j=1..4$ as shown in Table \ref{tab:ex_t_d}: trace $t_5$ relates to defect $d_1$; traces $t_1$ and $t_3$ relate to defect $d_2$; trace $t_4$ to defect $d_3$; and trace $t_2$ to defect $d_4$.

\begin{table}[htb]
\caption{Example: Relation between traces and defects}
\label{tab:ex_t_d}
\begin{ruledtabular}
\begin{tabular}{ll}

Defect
 & 
Trace
 \\
\hline

$d_{1}$
 & 
$t_{5}$
 \\

$d_{2}$
 & 
$t_{1},t_{3}$
 \\

$d_{3}$
 & 
$t_{4}$
 \\

$d_{4}$
 & 
$t_{2}$
 \\

\end{tabular}
\end{ruledtabular}
\end{table}

Suppose that we calculate distances between traces using some 
distance measure. The distances between (a potentially new) trace $t$ and each trace in $T=\{t_1,\ldots,t_5\}$ and the defects' ranks obtained using these hypothetical calculations are given in Table \ref{tab:ex_rank}.  Trace $t_{2}$ is the closest to $t$, hence $d_{4}$ (to which
$t_{2}$ is related) gets ranking number 1. Traces $t_{1}$ and $t_{4}$ have the same distance to $t$, therefore, $d_{2}$ and $d_{3}$ get the same rank. Based on the modified competition ranking schema algorithm we leave a gap before the set of items with the same rank and assign rank 3 to both traces. Traces $t_{3}$ and $t_{5}$
also have the same distance to $t$; however $t_{3}$ should be ignored since it relates to the already ranked defect $d_{2}$. This assigns rank 4 to $d_{1}$. The resulting sets of top $X$ traces for different values of
$X$ are shown in Table \ref{tab:ex_top}.

\begin{table}[htb]
\caption{Example: Traces sorted by distance and ranked}
\label{tab:ex_rank}
\begin{ruledtabular}
\begin{tabular}{lrrr}

$t_{i}$
 & 
Distance between 
 & 
Class (defect ID) 
 & 
Rank
 \\

 & 
$t$ and $t_i$
 & 
of trace $t_i$
 & 

 \\
\hline

$t_{2}$
 & 
0
 & 
$d_{4}$
 & 
1
 \\

$t_{1}$
 & 
7
 & 
$d_{2}$
 & 
3
 \\

$t_{4}$
 & 
7
 & 
$d_{3}$
 & 
3
 \\

$t_{3}$
 & 
9
 & 
$d_{2}$
 & 
--
 \\

$t_{5}$
 & 
9
 & 
$d_{1}$
 & 
4
 \\

\end{tabular}
\end{ruledtabular}
\end{table}

\begin{table}[htb]
\caption{Example: Top 1-4 defects}
\label{tab:ex_top}
\begin{ruledtabular}
\begin{tabular}{ll}
Top $X$
 & 
Set of defects in Top $X$
 \\
\hline

Top 1
 & 
$d_{4}$
 \\

Top 2
 & 
$d_{4}$
 \\

Top 3
 & 
$d_{4}$, $d_{2 }$,
$d_{3}$
 \\

Top 4
 & 
$d_{4}$, $d_{2 }$,
$d_{3}$, $d_{1}$
 \\

\end{tabular}
\end{ruledtabular}
\end{table}

\subsection{Traces ranking algorithm: efficiency}\label{sec:3.3}

The number of operations $C$ needed by the ranking algorithm is given by
\begin{eqnarray}\label{eq:complexity}
  C &=& \underbrace {{\tilde{c}_1}O(|M||T|)}_{{\text{Step 1}}} + \underbrace {{\tilde{c}_2}O(|T|\log |T|)}_{{\text{Step 2}}} + \underbrace {{\tilde{c}_3}O(|T|)}_{{\text{Step 3}}}  \nonumber \\
&+& \underbrace {{\tilde{c}_4}O(|T|)}_{{\text{Step 4}}} + \underbrace {{\tilde{c}_5}O(|T|)}_{{\text{Step 5}}} + \underbrace {{\tilde{c}_6}O(1)}_{{\text{Step 6}}}  \nonumber \\
&\mathop  \approx \limits^{\begin{subarray}{c} 
  |T| \to \infty,  \\ 
  \tilde{c}_1 \gg \{\tilde{c}_3,\tilde{c}_4,\tilde{c}_5\}
\end{subarray}}&
 \underbrace {{\tilde{c}_1}O(|M||T|)}_{{\text{Step 1}}} + \underbrace {{\tilde{c}_2}O(|T|\log |T|)}_{{\text{Step 2}}},  
\end{eqnarray}
where $\tilde{c}_{i}$ is a constant number of operations associated
with $i$-th step, and $|\cdot{}|$ represents the number of elements in a given set. In practice, the coefficients $\tilde{c}_3$, $\tilde{c}_4$ and $\tilde{c}_5$ are of much smaller order than $\tilde{c}_1$ and hence terms corresponding to Steps 3, 4 and 5 do not contribute significantly to $C$. The pairwise distance calculation via Equation~(\ref{eq:d_all})  requires $O(|M|)$ operations. Therefore, calculating all distances between traces (Step 1) requires
$O(|M||T|)$ operations, so for fixed $|M|$ the number of operations grows linearly with $|T|$. The average sorting algorithm, required by Step 2 (sorting of traces by their distance to trace $t$), needs $O(|T|\log|T|)$ operations \cite{cormen_introduction_2009}. Usually, ${\tilde{c}_1} \gg {\tilde{c}_2}$; this implies that a user may expect to see a linear relation between $C$ and $|T|$ (even for large $|T|$), in spite of the loglinear complexity of the second term in Equation~(\ref{eq:complexity}).

The amount of storage needed for entropy-based fingerprints data (used
by Equation~(\ref{eq:d_all})) is proportional to
\begin{eqnarray}
\underbrace {\phi |M||T|}_a + \underbrace {\phi |M|}_b = \phi |M|(|T| + 1),
\end{eqnarray}
where $\phi$ is the number of bytes needed to store a single fingerprint value. Term $a$ is the amount of storage needed for entropy-based fingerprints for all
traces in $T$, and term $b $ is the amount of storage needed for the values of $\max\left\{ {{H_{{E_k}}}\left[ {\alpha (t;{l_k},{\tilde{c}_k});{q_k}} \right]}
\right\}$ from Equation~(\ref{eq:d_all}). Assuming that $|M|$ remains constant, the data size grows linearly with $|T|$. Now we will compare the complexity of our algorithm with an existing technique.

\subsubsection{Comparison with existing algorithm}\label{sec:3.3.1}
Note that a general approach for selecting a subset of closest traces to a given trace can be summarized as follows: (a) calculate the distance between $t$ and each trace in $T$; and (b) select a subset of closest traces. The principal difference between various approaches lies in the technique for calculating the distance between pairs of traces.

Suppose that the distance between a pair of traces is calculated using the Levenshtein distance\footnote{The Levenshtein distance between two strings is given by the minimum number of operations (insertion, deletion, and substitution) needed to transform one string into another.} \cite{levenshtein_bin_1966}. This distance can be calculated using the difference algorithm \cite{myers_diff_1986}. The worst case complexity for comparing a pair of strings using this algorithm is $O(ND_L)$ \cite{myers_diff_1986}, where $N$ is the combined length of a pair strings and $D_L$ is the Levenshtein distance between these two strings. Therefore, the complexity of step (a) using this technique is $O(ND_L)$. 

Let us compare $O(ND_L)$ with the complexity of calculating the distance using the entropy based algorithm, namely $O(|M|)$. The former depends linearly on the length of the string representation of a trace ranging between $10^0$ and $10^8$ ``characters'' \cite{miranskyy_sift:_2008}. However, when strings are completely different, $D_L = N$ and $O(ND_L) \rightarrow O(N^2)$, implying quadratic dependency on the trace length. Conversely, $|M|$ (representing the quantity of scalar entropy values) should range between $10^0$ and $10^2$.  The computations are independent of trace size and require four mathematical function calls per scalar (see Equation~(\ref{eq:d_all})) when efficiently implemented on modern hardware.  Therefore, our algorithm requires several orders of magnitude fewer operations than the existing algorithms.

In order to implement the Levenshtein distance algorithm we must preserve the original traces, requiring storage space for $10^0$ to $10^8$ ``characters'' of each trace \cite{miranskyy_sift:_2008}. An entropy based fingerprint, on the other hand, will need to store $|M|$ real numbers ($|M|$ ranging from $10^0$ to $10^2$). This represents a reduction of several orders of magnitude in storage requirements.

\section{Validation case study}\label{sec:4}

Our hypothesis is that the predictive classification power will vary with changes in $E$, $l$, $c$, and $q$. In order to study the classification power of ${H_E}\left[ {\alpha(t;l,c);q} \right]$ we will analyze Cartesian products of the following sets of variables:

\begin{enumerate}
	\item $E \in (S,L,R,T)$,
	\item $l \in (1,2, \ldots ,7)$,
	\item $q \in (0,{10^{ - 5}},{10^{ - 4}}, \ldots,{10^1},{10^2})$,
	\item $c \in (F,FT,FTD)$.
\end{enumerate}

We denote by $\Lambda$ the complete set of parameters obtained by the Cartesian
product of the above sets.

For our validation case study we use the Siemens test suite, developed by Hutchins et al. \cite{hutchins_experiments_1994}.
This suite was further augmented and made available to the public at the Software-artifact
Infrastructure Repository \cite{do_supporting_2005, _siemens_????}. This software suite has been used in many 
defect analysis studies over the last decade (see \cite{jones_empirical_2005, murtaza_f007:_2010} for literature review) and hence provides an example on which to test our algorithm.

The Siemens suite \cite{hutchins_experiments_1994} contains seven programs. Each program has one original version and a number of faulty versions. Hutchins et al. \cite{hutchins_experiments_1994} created these faulty versions by making a single manual source code alteration to the original version. A fault can span multiple lines of source code and multiple
functions. Every such fault was created to cause logical rather than execution errors.  Thus, the faulty programs in the suite do not terminate prematurely after the execution of faulty code, they simply return incorrect output. Each program comes with a collection of test cases,
applicable to all faulty versions and the original program. A fault can
be identified if the output of a test case on the original version
differs from the output of the same test case on a faulty version of
the program. 

In this study, we experimented with the largest program (``Replace'') of the Siemens suite. It has 517 lines of code, 21 functions, and 31 different faulty versions. There were 5542 test cases shared across all the versions.
Out of these $31\times{}5542$ test cases, 4266 ($\approx 2.5 \%$ of the total number of test cases) caused a program failure when exposed to the faulty program, i.e., were able to catch a defect. The remaining test cases were
probably unrelated to the 31 defects. The traces for failed test cases
were collected using a tool called Etrace \cite{_etrace_????}.  The tool captures
sequences of function-calls for a particular software execution such as
the one shown in Figure~\ref{fig:trc_ex}. In other words, we collected 4266 function-call level failed traces for 31 faults (faulty versions) of the
``Replace'' program\footnote{ The ``Replace'' program had 32 faults,
but the tool ``Etrace'' was unable to capture the traces of segmentation fault in one of the faulty versions of the ``Replace'' program. This problem was also reported by other researchers \cite{jones_empirical_2005}.}.

\begin{figure}[htb]
	\includegraphics[width=250pt]{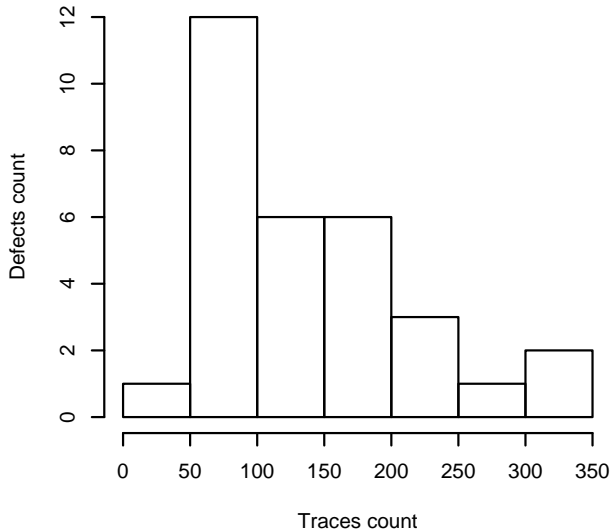}
\caption{Distribution of the number of traces per defect(version)}
\label{fig:trc_distr}
\end{figure}

\begin{figure}[htb]
\includegraphics[width=250pt]{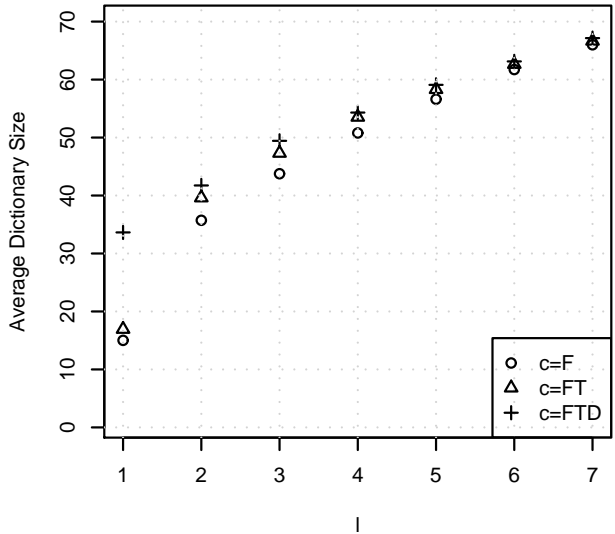}
\caption{Dictionary size for various values of $l$ and $c$.}
\label{fig:dic_sz}
\end{figure}

The distribution of the number of traces mapped to a particular defect
(version) is given in Figure~\ref{fig:trc_distr}. Descriptive statistics of trace length are given in Table~\ref{tab:trc_length_stat}. The length ranges between 11 and 101400 records per trace; with an average length of 623 records per trace. Average dictionary sizes for various values of $c$ are given in Figure~\ref{fig:dic_sz}. Note that as $l$ gets larger the dictionary sizes for all $c$ start to converge.

\begin{table}[htb]
\caption{Descriptive statistics of length of traces}
\label{tab:trc_length_stat}
\begin{ruledtabular}
\begin{tabular}{rrrrrr}
Min.
 & 
1$^{st}$ Qu.
 & 
Median
 & 
Mean
 & 
3$^{rd}$ Qu.
 & 
Max.
 \\
\hline

11
 & 
218
 & 
380
 & 
623.3
 & 
678
 & 
101400
 \\
\end{tabular}
\end{ruledtabular}

\end{table}

All of the traces contain at least one common function. Therefore, we
skip the pre-filtering step. Note that direct comparison with existing
trace comparison techniques is not possible since 1) the authors
focus on identification of faulty functions \cite{jones_empirical_2005, murtaza_f007:_2010} instead of
identification of defect IDs and 2) the authors \cite{jones_empirical_2005} analyze a complete set
of programs in the Siemens suite while we focus only on one program
(Replace).

The case study is divided as follows:  the individual classification power of each ${H_E}\left[{\alpha (t;l,c);q} \right]$ is analyzed in Section~\ref{sec:4.1},  while Section~\ref{sec:4.2} analyzes the classification power of the complete set of entropies. Timing analysis of the algorithm is given in Section~\ref{sec:timing}, and threats to validity are discussed in Section~\ref{sec:validity}.

\subsection{Analysis of individual entropies}\label{sec:4.1}

Analysis of the classification power of individual entropies is
performed using 10-fold cross-validation. In this approach the sample set is partitioned into 10 disjoint subsets of equal size. A single subset is used as validation data for testing and the remaining nine subsets are used as training data. The process is repeated tenfold (ten times) using each data subset as the validation data just once. The results of the validation from each fold are averaged to produce a single estimate. This 10-fold repetition guards against the sampling bias which can be introduced through the use of the more traditional 1-fold validation in which 70\% of the data is used for training and the remaining 30\% is used for testing. The validation process is designed as follows:

\begin{enumerate}
	\item Randomly partition 4266 traces into 10 bins
	\item For each set of parameters ${E,l,c,q}$
	\begin{enumerate}
		\item For each bin
			\begin{enumerate}
				\item Tag traces in a given bin as a validating set of data and traces in the remaining nine bins as a training set
				\item For each trace $t$ in the validating set calculate the rank of $t$'s class (defect ID) in the training set using the algorithm in Section~\ref{sec:3.2}\footnote{ Technically, in order to identify the true ranking one must tweak Step 6 of the algorithm and return a vector of 2-tuples [class, rank].
} with Equation~(\ref{eq:d_single}) as the measure of distance and with the set of parameters ${E,l,c,q}$
			\end{enumerate}
	\item Compute summary statistics about ranks of the ``true'' classes and store this data for further analysis
	\end{enumerate}
\end{enumerate}

\begin{figure}[htbp]
\includegraphics[width=250pt]{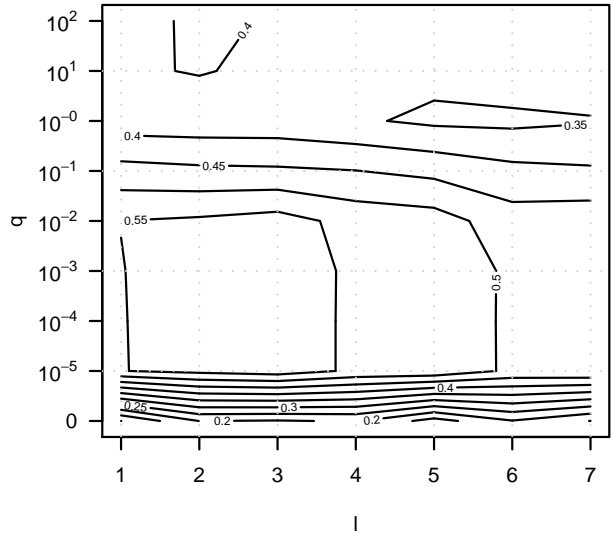}
\caption{Interpolated average fractions of correctly classified
traces in the Top~5 (based on 10-fold cross validation) for $E=L$, $l=3$,
$q=10^{-5}$, and $c=FTD$ for different values of $l$ and $q$.}
\label{fig:contour_l_q}
\end{figure}

\begin{figure}[htbp]
\includegraphics[width=250pt]{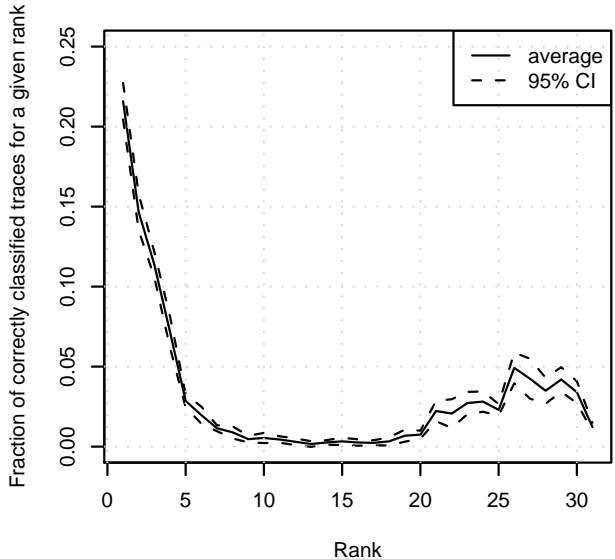}
\caption{Fraction of correctly classified traces in the Top~5 for $E=L$
and $c=FTD$. The solid line shows the average fraction of correctly
classified traces in 10 folds; the dotted line shows the pointwise $95\%$
confidence interval (95\% CI) of the average.}
\label{fig:best_rank_details}
\end{figure}

Our findings show that the best results are obtained for $H$ with $E \in (L,R,T)$, $l=3$, $q \in ({10^{ - 5}},{10^{ -4}})$, and $c = FTD$. Based on 10-fold cross validation, the entropies with these parameters were able to correctly classify $\approx 21.6\% \pm 1.1\%$\footnote{ $95\%$ confidence interval, calculated as $ \pm q(0.975,9)/\sqrt {10}  \times {\text{standard deviation}}$, where $q(x,df)$ represents quantile function of the $t$-distribution, $x$ is the probability, and $df$ is the degrees of
freedom.} 
of the Top 1 defects and $\approx 57.6 \% \pm 1.5\% $ of the Top 5 defects (see Table~\ref{tab:class_power} and Figure~\ref{fig:contour_l_q}). Based on the standard deviation data in Table~\ref{tab:class_power}, all six entropies show robust results. However, the results become slightly more volatile for high ranks (see Figure~\ref{fig:best_rank_details}). We now analyze these findings in details.

\begingroup
\squeezetable
\begin{table*}[htbp]

\caption{ Fraction of correctly classified traces in Top $X$ for 1)
${H_E}\left[ {\alpha (t;l,c);q} \right]$
with $E \in (L,R,T)$, $q \in (10^{-5}, 10^{-4})$, $l=3$, and $c=FTD$, and 2)  set of entropies $\Lambda$; based on 10-fold cross validation. The average fraction of correctly classified traces in 10 folds is denoted by ``Avg.''; plus-minus $95\%$ confidence
interval is denoted by ``95\% CI''.}
\label{tab:class_power}

\begin{ruledtabular}
\begin{tabular}{
lrrrrrrrrrrrrrr
}

& \multicolumn{4}{|c}{
 $E$=$L$
} & \multicolumn{4}{|c}{
 $E$=$R$
} & \multicolumn{4}{|c}{
 $E$=$T$
} & \multicolumn{2}{|c}{
 $\Lambda{}$
} \\
\cline{2-13} 
Top $X$
&  \multicolumn{2}{|c}{
 $q$=10$^{-4}$
} & \multicolumn{2}{|c}{
 $q$=10$^{-5}$
} &  \multicolumn{2}{|c}{
 $q$=10$^{-4}$
} & \multicolumn{2}{|c}{
 $q$=10$^{-5}$
} &  \multicolumn{2}{|c}{
 $q$=10$^{-4}$
} & \multicolumn{2}{|c}{
 $q$=10$^{-5}$
}  & \multicolumn{2}{|c}{

} \\
\cline{2-15} 
 & 
\multicolumn{1}{|c}{
 Avg.
}
&
\multicolumn{1}{|c}{
 95\% CI
}
 & 
\multicolumn{1}{|c}{
 Avg.
}
&
\multicolumn{1}{|c}{
 95\% CI
}
 & 
\multicolumn{1}{|c}{
 Avg.
}
&
\multicolumn{1}{|c}{
 95\% CI
}
 & 
\multicolumn{1}{|c}{
 Avg.
}
&
\multicolumn{1}{|c}{
 95\% CI
}
 & 
\multicolumn{1}{|c}{
 Avg.
}
&
\multicolumn{1}{|c}{
 95\% CI
}
 & 
\multicolumn{1}{|c}{
 Avg.
}
&
\multicolumn{1}{|c}{
 95\% CI
}
 & 
\multicolumn{1}{|c}{
 Avg.
}
&
\multicolumn{1}{|c}{
 95\% CI
}

 \\
 \hline

 1
 & 
 0.2159
 & 
 0.0114
 & 
 0.2159
 & 
 0.0114
 & 
 0.2159
 & 
 0.0114
 & 
 0.2159
 & 
 0.0114
 & 
 0.2159
 & 
 0.0114
 & 
 0.2159
 & 
 0.0114
 & 
 0.2982
 & 
 0.0160
 \\

 2
 & 
 0.3624
 & 
 0.0180
 & 
 0.3624
 & 
 0.0180
 & 
 0.3624
 & 
 0.0180
 & 
 0.3624
 & 
 0.0180
 & 
 0.3621
 & 
 0.0180
 & 
 0.3621
 & 
 0.0180
 & 
 0.4669
 & 
 0.0181
 \\

 3
 & 
 0.4761
 & 
 0.0219
 & 
 0.4761
 & 
 0.0219
 & 
 0.4761
 & 
 0.0219
 & 
 0.4761
 & 
 0.0219
 & 
 0.4761
 & 
 0.0219
 & 
 0.4761
 & 
 0.0219
 & 
 0.5769
 & 
 0.0194
 \\

 4
 & 
 0.5478
 & 
 0.0164
 & 
 0.5478
 & 
 0.0164
 & 
 0.5478
 & 
 0.0164
 & 
 0.5478
 & 
 0.0164
 & 
 0.5478
 & 
 0.0165
 & 
 0.5478
 & 
 0.0165
 & 
 0.6020
 & 
 0.0178
 \\

 5
 & 
 0.5764
 & 
 0.0149
 & 
 0.5764
 & 
 0.0149
 & 
 0.5764
 & 
 0.0149
 & 
 0.5764
 & 
 0.0149
 & 
 0.5766
 & 
 0.0151
 & 
 0.5766
 & 
 0.0151
 & 
 0.6214
 & 
 0.0183
 \\

 6
 & 
 0.5961
 & 
 0.0174
 & 
 0.5961
 & 
 0.0174
 & 
 0.5961
 & 
 0.0174
 & 
 0.5961
 & 
 0.0174
 & 
 0.5956
 & 
 0.0177
 & 
 0.5956
 & 
 0.0177
 & 
 0.6291
 & 
 0.0176
 \\

 7
 & 
 0.6076
 & 
 0.0174
 & 
 0.6076
 & 
 0.0174
 & 
 0.6076
 & 
 0.0174
 & 
 0.6076
 & 
 0.0174
 & 
 0.6076
 & 
 0.0173
 & 
 0.6076
 & 
 0.0173
 & 
 0.6352
 & 
 0.0177
 \\

 8
 & 
 0.6165
 & 
 0.0166
 & 
 0.6165
 & 
 0.0166
 & 
 0.6165
 & 
 0.0166
 & 
 0.6165
 & 
 0.0166
 & 
 0.6169
 & 
 0.0165
 & 
 0.6169
 & 
 0.0165
 & 
 0.6399
 & 
 0.0180
 \\

 9
 & 
 0.6212
 & 
 0.0162
 & 
 0.6212
 & 
 0.0162
 & 
 0.6212
 & 
 0.0162
 & 
 0.6212
 & 
 0.0162
 & 
 0.6228
 & 
 0.0158
 & 
 0.6228
 & 
 0.0158
 & 
 0.6444
 & 
 0.0177
 \\

 10
 & 
 0.6266
 & 
 0.0167
 & 
 0.6266
 & 
 0.0167
 & 
 0.6266
 & 
 0.0167
 & 
 0.6266
 & 
 0.0167
 & 
 0.6280
 & 
 0.0163
 & 
 0.6280
 & 
 0.0162
 & 
 0.6467
 & 
 0.0180
 \\

 11
 & 
 0.6310
 & 
 0.0172
 & 
 0.6310
 & 
 0.0172
 & 
 0.6310
 & 
 0.0172
 & 
 0.6310
 & 
 0.0172
 & 
 0.6331
 & 
 0.0167
 & 
 0.6331
 & 
 0.0168
 & 
 0.6481
 & 
 0.0179
 \\

 12
 & 
 0.6341
 & 
 0.0177
 & 
 0.6341
 & 
 0.0177
 & 
 0.6341
 & 
 0.0177
 & 
 0.6341
 & 
 0.0177
 & 
 0.6357
 & 
 0.0175
 & 
 0.6357
 & 
 0.0175
 & 
 0.6488
 & 
 0.0177
 \\

 13
 & 
 0.6357
 & 
 0.0170
 & 
 0.6357
 & 
 0.0170
 & 
 0.6357
 & 
 0.0170
 & 
 0.6357
 & 
 0.0170
 & 
 0.6381
 & 
 0.0169
 & 
 0.6381
 & 
 0.0170
 & 
 0.6491
 & 
 0.0176
 \\

 14
 & 
 0.6383
 & 
 0.0173
 & 
 0.6383
 & 
 0.0173
 & 
 0.6383
 & 
 0.0173
 & 
 0.6383
 & 
 0.0173
 & 
 0.6406
 & 
 0.0170
 & 
 0.6406
 & 
 0.0169
 & 
 0.6493
 & 
 0.0176
 \\

 15
 & 
 0.6416
 & 
 0.0164
 & 
 0.6416
 & 
 0.0164
 & 
 0.6420
 & 
 0.0162
 & 
 0.6416
 & 
 0.0164
 & 
 0.6432
 & 
 0.0162
 & 
 0.6432
 & 
 0.0162
 & 
 0.6500
 & 
 0.0174
 \\

 16
 & 
 0.6441
 & 
 0.0167
 & 
 0.6441
 & 
 0.0167
 & 
 0.6441
 & 
 0.0167
 & 
 0.6441
 & 
 0.0167
 & 
 0.6453
 & 
 0.0166
 & 
 0.6453
 & 
 0.0166
 & 
 0.6505
 & 
 0.0173
 \\

 17
 & 
 0.6463
 & 
 0.0157
 & 
 0.6465
 & 
 0.0157
 & 
 0.6463
 & 
 0.0157
 & 
 0.6465
 & 
 0.0157
 & 
 0.6474
 & 
 0.0156
 & 
 0.6477
 & 
 0.0155
 & 
 0.6516
 & 
 0.0168
 \\

 18
 & 
 0.6495
 & 
 0.0147
 & 
 0.6498
 & 
 0.0146
 & 
 0.6495
 & 
 0.0147
 & 
 0.6498
 & 
 0.0146
 & 
 0.6509
 & 
 0.0144
 & 
 0.6512
 & 
 0.0144
 & 
 0.6542
 & 
 0.0161
 \\

 19
 & 
 0.6563
 & 
 0.0150
 & 
 0.6566
 & 
 0.0149
 & 
 0.6563
 & 
 0.0150
 & 
 0.6566
 & 
 0.0149
 & 
 0.6577
 & 
 0.0147
 & 
 0.6580
 & 
 0.0147
 & 
 0.6570
 & 
 0.0156
 \\

 20
 & 
 0.6641
 & 
 0.0163
 & 
 0.6641
 & 
 0.0163
 & 
 0.6641
 & 
 0.0163
 & 
 0.6641
 & 
 0.0163
 & 
 0.6655
 & 
 0.0161
 & 
 0.6655
 & 
 0.0160
 & 
 0.6641
 & 
 0.0159
 \\

 21
 & 
 0.6863
 & 
 0.0142
 & 
 0.6863
 & 
 0.0142
 & 
 0.6863
 & 
 0.0142
 & 
 0.6863
 & 
 0.0142
 & 
 0.6873
 & 
 0.0141
 & 
 0.6873
 & 
 0.0142
 & 
 0.6725
 & 
 0.0176
 \\

 22
 & 
 0.7070
 & 
 0.0106
 & 
 0.7070
 & 
 0.0106
 & 
 0.7070
 & 
 0.0106
 & 
 0.7070
 & 
 0.0106
 & 
 0.7079
 & 
 0.0104
 & 
 0.7079
 & 
 0.0104
 & 
 0.6971
 & 
 0.0171
 \\

 23
 & 
 0.7342
 & 
 0.0117
 & 
 0.7342
 & 
 0.0117
 & 
 0.7339
 & 
 0.0117
 & 
 0.7339
 & 
 0.0117
 & 
 0.7342
 & 
 0.0114
 & 
 0.7342
 & 
 0.0112
 & 
 0.7178
 & 
 0.0158
 \\

 24
 & 
 0.7623
 & 
 0.0092
 & 
 0.7623
 & 
 0.0092
 & 
 0.7623
 & 
 0.0092
 & 
 0.7623
 & 
 0.0092
 & 
 0.7628
 & 
 0.0089
 & 
 0.7628
 & 
 0.0088
 & 
 0.7527
 & 
 0.0124
 \\

 25
 & 
 0.7853
 & 
 0.0076
 & 
 0.7853
 & 
 0.0076
 & 
 0.7855
 & 
 0.0074
 & 
 0.7855
 & 
 0.0074
 & 
 0.7865
 & 
 0.0072
 & 
 0.7865
 & 
 0.0072
 & 
 0.7834
 & 
 0.0134
 \\

 26
 & 
 0.8345
 & 
 0.0082
 & 
 0.8345
 & 
 0.0082
 & 
 0.8347
 & 
 0.0077
 & 
 0.8347
 & 
 0.0077
 & 
 0.8352
 & 
 0.0082
 & 
 0.8352
 & 
 0.0082
 & 
 0.8141
 & 
 0.0078
 \\

 27
 & 
 0.8769
 & 
 0.0097
 & 
 0.8769
 & 
 0.0097
 & 
 0.8772
 & 
 0.0097
 & 
 0.8772
 & 
 0.0097
 & 
 0.8776
 & 
 0.0097
 & 
 0.8776
 & 
 0.0097
 & 
 0.8591
 & 
 0.0109
 \\

 28
 & 
 0.9119
 & 
 0.0082
 & 
 0.9119
 & 
 0.0082
 & 
 0.9119
 & 
 0.0082
 & 
 0.9119
 & 
 0.0082
 & 
 0.9142
 & 
 0.0083
 & 
 0.9142
 & 
 0.0083
 & 
 0.9140
 & 
 0.0126
 \\

 29
 & 
 0.9538
 & 
 0.0054
 & 
 0.9538
 & 
 0.0054
 & 
 0.9538
 & 
 0.0054
 & 
 0.9538
 & 
 0.0054
 & 
 0.9550
 & 
 0.0054
 & 
 0.9550
 & 
 0.0059
 & 
 0.9618
 & 
 0.0068
 \\

 30
 & 
 0.9878
 & 
 0.0027
 & 
 0.9878
 & 
 0.0027
 & 
 0.9878
 & 
 0.0027
 & 
 0.9878
 & 
 0.0027
 & 
 0.9887
 & 
 0.0027
 & 
 0.9887
 & 
 0.0027
 & 
 0.9988
 & 
 0.0014
 \\

 31
 & 
 1.0000
 & 
 0.0000
 & 
 1.0000
 & 
 0.0000
 & 
 1.0000
 & 
 0.0000
 & 
 1.0000
 & 
 0.0000
 & 
 1.0000
 & 
 0.0000
 & 
 1.0000
 & 
 0.0000
 & 
 1.0000
 & 
 0.0000
 \\
\end{tabular}
\end{ruledtabular}

\end{table*}
\endgroup

\begin{figure}[htbp]
\includegraphics[width=250pt]{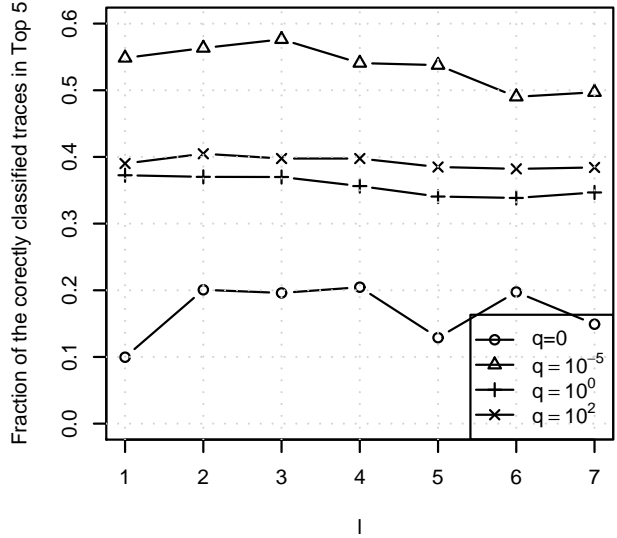}
\caption{The average fraction of traces correctly classified in the Top~5
for various values of $l$; $E = L$, $q \in (0,{10^{ - 5}},{10^{ 0 }},{10^2})$, $c=FTD$.}
\label{fig:best_l}
\end{figure}

The $3$-words ($l=3$) provide the best results based on the fraction of correctly classified traces in the Top 5 (see Figure~\ref{fig:best_l}), suggesting that chains of three events provide an optimal balance between the amount of information in a given $l$-word and the total number of words. As $l$ gets larger, the amount of data becomes insufficient to obtain a good estimate of the probabilities.

Examining the classification performance of $c$ (see Table~\ref{tab:class_power_FTD}) we see essentially no differences across the three levels of $c$ ($F,FT$, and $FTD$) considered here (with $E=L, l=3$, and $q\in\{10^{-4},10^{-5}\}$).  This is true across all values of $X$ in the percentage of correctly classified traces in the Top $X$.  This is somewhat surprising since $c=FTD$ contains more information than $c=FT$ which, in turn, contains more information than $c=F$.  This suggests that for this data set and parameters ($E=l$, $l=3$, and $q\in\{10^{-4},10^{-5}\}$) the function names contain the relevant classification information, the function type (entry or exit) and depth providing no additional relevant information.  Note that even though more time is needed to calculate the $FTD$-based entropies (since the dictionary of $FTD$s will contain twice as many entries as the dictionary of $FT$s) the comparison time remains the same (since the probabilities of $l$-words, $P$, map to a scalar value via the entropy function for all values of $c$).

\begin{table}[htbp]
\caption{Percent of correctly classified traces in Top $X$ for 
${H_E}\left[ {\alpha (t;l,c);q} \right]$, $E=L$, $l=3$, and $q=\{10^{-4},10^{-5}\}$.}
\label{tab:class_power_FTD}
\begin{ruledtabular}
\begin{tabular}{lrrr}

 Top $X$
 & 
 $c=F$ 
 & 
 $c=FT$
 & 
 $c=FTD$
 \\
\hline

 Top 1
 & 
 21.7\%
 & 
 20.9\%
 & 
 21.6\%
 \\

 Top 2
 & 
 37.2\%
 & 
 35.3\%
 & 
 36.2\%
 \\

 Top 3
 & 
 49.5\%
 & 
 46.7\%
 & 
 47.6\%
 \\

 Top 4
 & 
 54.0\%
 & 
 53.5\%
 & 
 54.8\%
 \\

 Top 5
 & 
 56.2\%
 & 
 56.6\%
 & 
 57.6\%
 \\
\end{tabular}
\end{ruledtabular}
\end{table}

\begin{figure}[htbp]
\includegraphics[width=250pt]{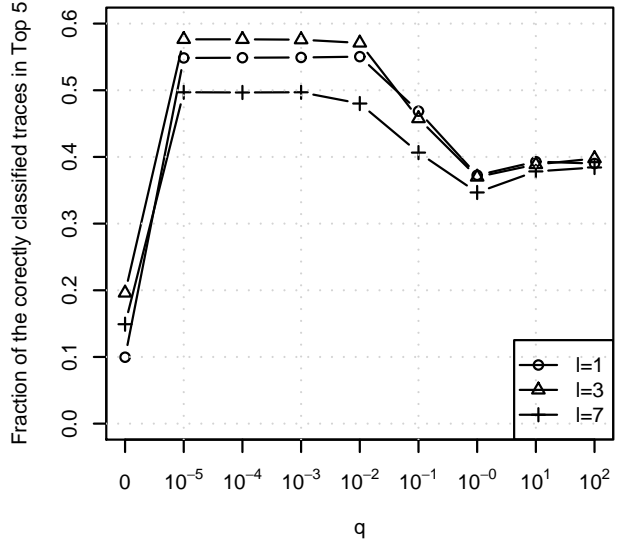}
\caption{Average fraction of correctly classified traces in the Top~5 for
various values of $q$; $E = L$, $l \in (1,3,7)$, $c~=~FTD$.}
\label{fig:best_q}
\end{figure}

Our findings show that the extended entropies outperform the Shannon
entropy\footnote{ We do not explicitly mention entropy values on the
figures. However, extended entropy values with $q=1$ correspond to values of the Shannon entropy.} for $q<1$ and $q>1$ (see Figure~\ref{fig:best_q}). However, performance of extended entropies with $q<1$ is significantly better than with $q>1$, suggesting that rare events are more important than frequent events for classification of defects in this dataset. The best results are obtained for $q=10^{-4}$ and $q=10^{-5}$.

It is interesting to note that classification performance is almost identical for $H$ with $E \in (L,R,T)$, $l$=3, $q \in ({10^{- 5}},{10^{ - 4}})$, and $c = FTD$. We believe that this fact can be explained as follows: the key
contribution to the ordering of similar traces (with similar
dictionaries) for entropies with $q \to 0$ is affected mainly by a function of probabilities of traces' events. This function is independent of $E$ and $q$ and depends only on $l$ and $c$, see Appendix \ref{sec:approx_of_d} for details.

\subsection{Analysis of the complete set of entropies}\label{sec:4.2}

Analysis of the classification power for the complete set of entropies
is performed using 10-fold cross-validation in a similar manner to the
process described in Section~\ref{sec:4.1}. However, instead of calculating
distances for each $H$ independently, we now calculate distances between traces by utilizing values of $H$ for all parameter sets in $\Lambda$ simultaneously. The validation process is designed as follows

\begin{enumerate}
	\item Randomly partition 4266 traces into 10 bins
\begin{enumerate}
	\item For each bin
\begin{enumerate}
	\item Tag traces in a given bin as a validating set of data and traces in
the remaining nine bins as a training set
	\item For each trace $t$ in the validating set calculate
the rank of $t$'s class (defect ID) in the training
set using the algorithm in Section~\ref{sec:3.2}  with Equation~(\ref{eq:d_all}) and all\footnote{ We had to exclude a subset of entropies with $E=L$, $q=10^2$ for all $l$ and $c$ from $\Lambda$. The
values of entropies obtained with these parameters are very large
($> 10^{100}$), which leads to numerical instability of Equation~(\ref{eq:d_all}). We keep just one of the various named $q = 1$ entropies to avoid redundancy.} 
4-tuples of parameters in $\Lambda$. (Based on the experiments described in Appendix~\ref{sec:p} we set $w=1$.)
\end{enumerate}
	\item Compute summary statistics about ranks of the ``true'' classes and store this data for further analysis
\end{enumerate}
\end{enumerate}

The results, in the two right-most columns of Table~\ref{tab:class_power}, show the increase of predictive power: in the case of the Top 1 the results improved from 21.6\% (for individual
entropies) to 29.8\% (for all entropies combined); for the Top~5 from
57.6\% to 62.1\%. However, the 503-fold increase in computational effort (the number of entropy fingerprints rise from 1 to 504)  yielded only an 8\% increase in the power to predict Top~5 matches. We leave the resulting balance between cost and benefit for each individual analyst to make.

\subsection{ Timing }\label{sec:timing}

We compared the theoretical efficiency of our algorithm with existing algorithms in Section~\ref{sec:3.3.1}. These findings may be validated against experimental timing data\footnote{Computations were performed on a computer with Intel Core 2 Duo E6320 CPU.}. We measured the time needed to compare a reference trace against a complete set of traces using difference \cite{myers_diff_1986} and entropy-based algorithms. The experiment is repeated three times for three reference traces: small (498 characters), medium (2339 characters), and large (20767 characters)\footnote{The reference traces were chosen arbitrarily.}. The results are given in Table~\ref{tab:timing}. As expected, the difference-based algorithm comparison time increases in proportion to the length of the reference trace, while the comparison time of the entropy-based algorithms remains constant across trace sizes. Moreover, the comparison time of entropy-based algorithms is several orders of magnitude faster than of the difference-based approach. Note that, even though the computational efforts for measuring distance based on the complete set of fingerprints increases by two orders of magnitude as compared to individual entropies, the results are obtained in less than a second. Therefore, from a practical perspective, we can still use a complete set of entropies for our analysis.

\begin{table}[htbp]
\caption{Timing results (in seconds) for comparison of a single reference trace against a set of traces using difference and entropy algorithms. }
\label{tab:timing}
\begin{ruledtabular}
\begin{tabular}{l|ccc}
   &   \multicolumn {3}{c} {Reference trace} \\
 Algorithm   &   Small & Medium  & Large \\
  &  (498 characters) & (2339 characters) & (20767 characters) \\
\hline  
 Difference & 2.7E1 & 4.2E1 & 1.9E2  \\
 Individual entropy  & 1.3E-4 & 1.3E-4 & 1.3E-4 \\
 Complete set of entropies & 6.8E-2 & 6.8E-2 & 6.8E-2 \\
\end{tabular}
\end{ruledtabular}
\end{table}

\subsection{ Threats to Validity }\label{sec:validity}
A number of tests are used to determine the quality of case studies. In this section we discuss four core tests: construct, internal, statistical, and external validity \cite{yin_case_2003}. The discussion highlights potential threats to validity of our case study and tactics that we used to mitigate the threats.

\textit{Construct validity}: to overcome potential construct validity issues we use two measures of classification performance (Top 1 and Top 5 correctly classified traces).
\textit{Internal validity}: to prevent data gathering issues all data collection was automated and a complete corpus of test cases was analyzed.
\textit{Statistical validity}: to prevent sampling bias, 10-fold cross validation was used.
\textit{External validity}: This case study shows that the method can be successfully applied to a particular, well-studied, data set.   Following the paradigm of the ``representative'' or ``typical'' case advanced in \cite{yin_case_2003}, this suggests that the method may also be useful in more situations.

\section{ Summary }\label{sec:5}

In this work we analyze the applicability of a technique which uses entropies to perform predictive
classification of traces related to software defects. Our validating
case study shows promising performance of extended entropies with
emphasis on rare events $\left( {q \in \left\{ 10^{-5}, 10^{-4} \right\}} \right)$. The events are based on triplets (3-words) of
``characters'' incorporating information about function name, depth of function call, and type of probe point ($c = FTD$).

In the future, we are planning to increase the number of datasets
under study, derive additional measures of distance (e.g., using tree
classification algorithms) and identify an optimal set of parameter combinations.

\appendix
\section{Selection of $w$ for Equation~(\ref{eq:d_all})}\label{sec:p}
We do not have a theoretical rationale for selection of the optimal value of $w$ in Equation~(\ref{eq:d_all}). When $|M|=1$, Equation~(\ref{eq:d_all}) simplifies to Equation~(\ref{eq:d_single}), becoming independent of $w$. However, in the general case of $|M|>1$, $w$ will affect performance of the distance metrics. In order to select an optimal value of $w$ we performed the analysis of the complete set of entropies discussed in Section~\ref{sec:4.2} for $w=1, 2, \ldots, 5$. Table~\ref{tab:p} gives the percentage of correctly classified traces in the Top~1 and the Top~5. The quality of classification does not change significantly with $w$. It degrades with increased $w$, but the small magnitude of this change makes us unwilling to consider it a result of the paper.

\begin{table}[htbp]
\caption{Percent of correctly classified traces in Top $X$ for the complete set of entropies $\Lambda$ and  $w=1, 2, \ldots, 5$.}
\label{tab:p}
\begin{ruledtabular}
\begin{tabular}{lrrrrrr}
 $w$ & 1 & 2 & 3 & 4 & 5 \\
\hline  
 Top 1 & 29.8\% & 29.7\% & 29.6\% & 29.3\% & 29.2\% \\
 Top 5 & 62.1\% & 61.5\% & 61.5\% & 61.3\% & 61.4\% \\
\end{tabular}
\end{ruledtabular}
\end{table}

\section{Approximation of Equation~(\ref{eq:d_single})}
\label{sec:approx_of_d}
We have observed that the classification power of the ${H_E}\left[ {\alpha (t;l,c);q} \right]$ metric is highest when $q \to 0$.
In order to explain this phenomenon we expand ${H_E}\left[ {\alpha(t;l,c);q} \right]$ (given in Equation~(\ref{eq:LHT})) in a Taylor series:

\begin{eqnarray}\label{eq:taylor}
  {H_L}\left[ {\alpha ({t_i};l,c);q} \right] &\mathop = \limits^{q \to 0}&  1 - \frac{1}
{{{n_i}}} 
+ q\left( \frac{{{A_i}}}{{n_i^2}} + \frac{{{n_i} - 1}}{{{n_i}}} \right)
+ O({q^2}), \nonumber \\
  {H_R}\left[ {\alpha ({t_i};l,c);q} \right] &\mathop  = \limits^{q \to 0}& {\log _2}({n_i})  \\
&+& q\left[ {\frac{{{A_i}}}
{{\ln (2){n_i}}} + {{\log }_2}({n_i})} \right] + O({q^2}), \text{ and} \nonumber \\
  {H_T}\left[ {\alpha ({t_i};l,c);q} \right] &\mathop  = \limits^{q \to 0}& {n_i} - 1 + q\left( {{A_i} + n - 1} \right) + O({q^2}), \nonumber 
\end{eqnarray}
where $A_i = \sum\limits_{k = 1}^{{n_i}} {\ln ({p_k}) }$.
By plugging (\ref{eq:taylor})  into Equation~(\ref{eq:d_single}) and assuming that for similar traces
$n \approx {n_i} \approx {n_j}$, we get:
\begin{eqnarray}\label{eq:d_approx}
  D({t_i},{t_j};L,q,l,c) &\approx& \frac{q}
{{{n^2}}}\left| {{A_i} - {A_j}} \right|, \nonumber \\
  D({t_i},{t_j};R,q,l,c) &\approx& \frac{q}
{{\ln (2)n}}\left| {{A_i} - {A_j}} \right|, \text{ and} \\
  D({t_i},{t_j};T,q,l,c) &\approx& q\left| {{A_i} - {A_j}} \right|. \nonumber
\end{eqnarray}

Equation~(\ref{eq:d_approx}) can be interpreted as follows. In the case in which $q \to 0$ and the dictionaries for each trace in the pair are similar, the key contribution to the measure of distance is coming from the $\sum_{k = 1}^{n_i} {\ln ({p_k})}$ term (which depends only on $l$ and $c$) making the rest of the variables irrelevant ($q$ and $n$ become parts of scaling factors). This can be highlighted by solving a system of equations to identify conditions that generate the same ordering for three traces ${t_i},{t_j},{t_k}$ for all
extended entropies (using approximations from (\ref{eq:d_approx})):

\begin{eqnarray}
  \left\{ \begin{gathered}
  \frac{q}
{{{n^2}}}\left| {{A_i} - {A_j}} \right| \leqslant \frac{q}
{{{n^2}}}\left| {{A_i} - {A_k}} \right|  \\
  \frac{q}
{{\ln (2)n}}\left| {{A_i} - {A_j}} \right| \leqslant \frac{q}
{{\ln (2)n}}\left| {{A_i} - {A_k}} \right|  \\
  q\left| {{A_i} - {A_j}} \right| \leqslant  q\left| {{A_i} - {A_k}} \right| \\ 
\end{gathered} 
\right. \Rightarrow  
\\
\Rightarrow \left| {{A_i} - {A_j}} \right| \leqslant \left| {{A_i} - {A_k}} \right|. \nonumber
\end{eqnarray}

In information theory $\ln(p_k)$ measures the ``surprise'' (in bits) in receiving symbol $k$ which is received with probability $p_k$. Thus $\sum_{k=1}^{n_i}{p_k \ln (p_k)}$ is the expected surprise or information (Shannon entropy). What about just $\sum_{k=1}^{n_i}{ \ln (p_k)}$? It scales with the total number of bits needed to specify each symbol.  This is related to the problem of simulating processes in the presence of rare events, see \cite{rubinshtein_optimization_1997} for details.  


\end{document}